\documentstyle[aps,prb,epsfig,float,twocolumn]{revtex}
\begin{document}
\twocolumn[\hsize\textwidth\columnwidth\hsize\csname
@twocolumnfalse\endcsname

\title{Anisotropic transport in the two-dimensional electron gas
in the presence of spin-orbit coupling}

\author{John Schliemann and Daniel Loss}

\address{Department of Physics and Astronomy, University of Basel, 
CH-4056 Basel, Switzerland}

\date{\today}

\maketitle

\begin{abstract}
In a two-dimensional electron gas as realized by a semiconductor
quantum well, the presence of spin-orbit coupling of both the Rashba and
Dresselhaus type leads to anisotropic dispersion relations and
Fermi contours. We study the effect of this anisotropy on the electrical
conductivity in the presence of fixed impurity scatterers. 
The conductivity also shows in general an anisotropy which
can be tuned by varying the Rashba coefficient. This effect provides
a method of detecting and investigating spin-orbit coupling by measuring
spin-unpolarized electrical currents in the diffusive regime. 
Our approach is based on an 
exact solution of the two-dimensional Boltzmann equation and provides
also a natural framework for investigating other transport effects
including the anomalous Hall effect. 
\end{abstract}
\vskip2pc]

\section{Introduction}

In the recent years the emerging field of spintronics 
\cite{Wolf01,Awschalom02} has generated an intense interest in effects of
spin-orbit interaction in low-dimensional semiconductor heterostructures.
For conduction band electrons in zinc-blende semiconductors the dominant
effects of spin-orbit interaction in low-dimensional geometry
can be described in terms of two effective contributions to the Hamiltonian. 
On the one hand there is the Rashba spin-orbit term \cite{Rashba60}
which is due to the inversion-asymmetry of the
confining potential and has the form
\begin{equation}
{\cal
H}_{R}=\frac{\alpha}{\hbar}\left(p_{x}\sigma^{y}-p_{y}\sigma^{x}\right),
\label{rashba}
\end{equation}
where $\vec p$ is the momentum of the electron confined in a
two-dimensional geometry, and $\vec\sigma$ the vector of Pauli
matrices. The coefficient $\alpha$ is tunable in strength by the
external gate perpendicular to the plane of the two-dimensional electron gas.
The other contribution is the Dresselhaus spin-orbit term which is present
in semiconductors lacking bulk inversion symmetry\cite{Dresselhaus55}. 
When restricted to a two-dimensional semiconductor nanostruture grown along 
the $[001]$ direction this coupling is of the form 
\cite{Dyakonov86,Bastard92}
\begin{equation}
{\cal H}_{D}=\frac{\beta}{\hbar}\left(p_{y}\sigma^{y}-p_{x}\sigma^{x}\right),
\label{dressel}
\end{equation}
where the coefficient $\beta$ is determined by the semiconductor
material and the geometry of the sample.

The interplay of these two types of spin-orbit coupling has been investigated
theoretically with respect to several physical phenomena
including the spin-splitting in zero magnetic field
\cite{Das90,Andrada97,Voskoboynikov01,Kainz03}, 
spin precession and re\-la\-xation
\cite{Kainz03,Dyakonov86,Bastard92,Averkiev99,Golub00,Kiselev00,Schliemann03,Saikin02,Lusakowski03,Winkler03}, contributions to magneto-oscillations 
\cite{Tarasenko02},
quantum interference corrections to the conductivity \cite{Pikus95,Miller03}, 
mesoscopic transport through quantum dots \cite{Aleiner01},
and issues of gate operations between quantum dot spin-qubits 
\cite{Stepanenko03}. In the present work we point out another effect occurring
in the presence of both Rashba and Dresselhaus spin-orbit coupling, namely
an anisotropy in the electrical conductivity for diffusive spin-unpolarized
charge transport in the presence of fixed impurity scatterers.

One of the most important concepts in the discussion of spin-orbit coupling
in semiconductors, and in the field of spintronics in general, is the 
spin field-effect transistor proposed by Datta and Das \cite{Datta90}.
This proposal uses the Rashba
spin-orbit coupling to perform controlled rotations of spins of
electrons passing through an FET-type device operating in the ballistic
transport regime. Ballistic transport is necessary 
to avoid randomization of the spin state by even spin-independent 
scatterers which would change the electron
momentum and therefore also the effective field provided by the
Rashba term in an uncontrolled way. The requirement of ballistic transport
has been so far one of the major obstacles toward the practical realization
of a spin FET. Recently an alternative scenario for a spin FET was proposed
which can also operate in the nonballistic regime \cite{Schliemann03}. This
proposal exploits the fact that if the Rashba coefficient is tuned via
external gates to be equal to the Dresselhaus coefficient,
$\alpha=\beta$, a new conserved quantity arises which prohibits the
randomization of the spin. As we shall see below, this particular point
$\alpha=\beta$ in parameter space will also be of special interest in our
present study. 

Our description of diffusive two-dimensional transport in the
presence of spin-orbit interaction and fixed impurities is based on an 
exact solution of the two-dimensional Boltzmann equation and provides
also a natural framework for investigating other transport effects
including the anomalous Hall effect. In particular, our scheme
of generating exact solutions to the transport equation can deal with
arbitrary dispersion relations and is not restricted to isotropic cases.

Another study on the possible influence of the  
Dresselhaus spin orbit coupling on the operation of the spin FET
was performed very recently by Lusakowski, Wrobel, and Dietl 
\cite{Lusakowski03}. These
investigations are restricted to the ballistic regime, but take into account,
in addition to the Hamiltonian (\ref{dressel}), also contributions to the
Dresselhaus term being trilinear in the momentum. In the present study we will
mostly neglect these trilinear contributions, but discuss their possible
influence briefly in section \ref{trilinear}. In another very recent
preprint R. Winkler also studied spin transport in the presence
of spin-orbit interaction steming from both structure inversion asymmetry
and bulk inversion asymmetry \cite{Winkler03}. 
In particular, in Refs.~\cite{Lusakowski03,Winkler03} 
the possibilities arising from 
different growth directions for the two-dimensional electron system
are explored. Yet another recent work dealing with transport
in the presence of spin-orbit coupling was performed by Mishchenko
and Halperin who derived the equations of motion for the
single electron density matrix in Wigner representation in a two-dimensional
free electron gas \cite{Mishchenko03}.
The authors applied their results to the dynamic 
conductivity of the system taking into account, however, the Rashba term 
only such that no anisotropy in conductivity occurred.
Finally we mention a very recent work by Ganichev {\it et al.} 
who present an experimental method to distinguish the effects of Rashba and
Dresselhaus spin-orbit coupling using the spin galvanic effect
\cite{Ganichev03}.

This paper is organized as follows. In section \ref{dispersion}
we review the dispersion relations and eigenstates of free electrons
confined in two dimensions in the presence of spin orbit coupling 
of both the Rashba and the Dresselhaus type, and we present 
results for the Fermi contours. In section \ref{Boltzmann} 
we present a scheme to generate exact solutions to the two-dimensional
Boltzmann equation that underlies our present study. This
approximation-free solution to the semiclassical transport equation
is then applied in section
\ref{results} to the case of free electrons being subject
to spin-orbit interaction of the above type. We close with a summary
and discussion of the results in section \ref{summary}.


\section{Dispersion relations, eigenstates, and Fermi contours}
\label{dispersion}

We consider the single-particle Hamiltonian for a two-dimensional electron
system
\begin{equation}
{\cal H}=\frac{\vec p^{2}}{2m}+{\cal H}_{R}+{\cal H}_{D}
\end{equation}
where $m$ is an effective band mass. The eigenenergies are given by
\begin{equation}
\varepsilon_{\pm}\left(\vec k\right)
=\frac{\hbar^{2}k^{2}}{2m}
\pm\sqrt{\left(\alpha k_{y}+\beta k_{x}\right)^{2}
+\left(\alpha k_{x}+\beta k_{y}\right)^{2}}
\label{dispersion1}
\end{equation}
with eigenstates
\begin{equation}
\langle\vec r|\vec k,\pm\rangle=
\frac{e^{i\vec k\cdot\vec r}}{\sqrt{A}}\frac{1}{\sqrt{2}}
\left(
\begin{array}{c}
1 \\
\pm e^{i\chi(\vec k)}
\end{array}
\right)
\label{eigenstate}
\end{equation}
where $A$ is the area of the system and
\begin{equation}
\chi(\vec k)=\arg(-\alpha k_{y}-\beta k_{x}+i(\alpha k_{x}+\beta k_{y}))\,.
\label{chi}
\end{equation}
The semiclassical particle velocities are given by
\begin{eqnarray}
\vec v_{\pm}(\vec k) & = &
\frac{\partial\varepsilon_{\pm}(\vec k)}{\hbar\partial\vec k}
=\frac{\hbar\vec k}{m}\nonumber\\
 & & \pm\frac{\left(\alpha^{2}+\beta^{2}\right)\vec k
+2\alpha\beta\left(\tilde\sigma^{x}\vec k\right)}
{\hbar\sqrt{\left(\alpha^{2}+\beta^{2}\right)k^{2}
+2\alpha\beta\left(\vec k^{T}\tilde\sigma^{x}\vec k\right)}}
\end{eqnarray}
where $\tilde\sigma^{x}$ is a usual Pauli matrix acting on the vector 
$\vec k$. As a consistency check let us consider the quantum mechanical
velocity operator
\begin{equation}
\dot{\vec r}=\frac{i}{\hbar}\left[{\cal H},\vec r\right]\,.
\end{equation}
Using the above expressions for the eigenstates it is straightforward to
show that its matrix elements are given by
\begin{equation}
\langle\vec k,\pm|\dot{\vec r}|\vec k',\pm\rangle=
\delta_{\vec k,\vec k'}\vec v_{\pm}(\vec k)\,,
\end{equation}
i.e. the semiclassical velocities $\vec v_{\pm}(\vec k)$
are, as usual, the diagonal elements
of the velocity operator.

Parametrizing wave vectors as $\vec k=k(\cos\varphi,\sin\varphi)$ one obtains
for positive Fermi energy $\varepsilon_{f}$ the following
parametrization of the Fermi contours:
\begin{eqnarray}
& & k^{f}_{\pm}\left(\varphi;\varepsilon_{f}\right)
=\mp\sqrt{\left(\frac{m}{\hbar^{2}}\right)^{2}
\left(\alpha^{2}+\beta^{2}+2\alpha\beta\sin\left(2\varphi\right)\right)}
\nonumber\\
 & & +\sqrt{\frac{2m}{\hbar^{2}}\varepsilon_{f}
+\left(\frac{m}{\hbar^{2}}\right)^{2}
\left(\alpha^{2}+\beta^{2}+2\alpha\beta\sin\left(2\varphi\right)\right)}\,.
\label{fermimod1}
\end{eqnarray}
Here the double sign corresponds to the above two dispersion branches, and the
Fermi wave vector is given by
\begin{equation}
\vec k^{f}_{\pm}\left(\varphi;\varepsilon_{f}\right)
=k^{f}_{\pm}\left(\varphi;\varepsilon_{f}\right)(\cos\varphi,\sin\varphi)\,.
\label{fermivec}
\end{equation}
At negative Fermi energies the Fermi contours can become somewhat more
complicated. This case corresponds to rather low electron densities and
shall not be considered here further. In the following the Fermi energy is 
always assumed to be positive. From Eq.~(\ref{fermimod1}) one finds the
electron density $n$ as
\begin{eqnarray}
n & = & \frac{1}{(2\pi)^{2}}\sum_{\mu=\pm}
\int d\varphi\frac{1}{2}
\left(k^{f}_{\mu}\left(\varphi;\varepsilon_{f}\right)\right)^{2}
\nonumber\\
 & = & \frac{1}{2\pi}\left(\frac{2m}{\hbar^{2}}\varepsilon_{f}
+2\left(\frac{m}{\hbar^{2}}\right)^{2}
\left(\alpha^{2}+\beta^{2}\right)\right)\,.
\label{density}
\end{eqnarray}

If $\alpha=0$ or $\beta=0$ the dispersions are isotropic and Fermi contours
are concentric circles. For $\alpha\neq 0\neq\beta$ the Fermi contours
are anisotropic which, as we shall see below, leads in general 
to anisotropic transport properties.
Note that the dispersion relations and Fermi contours are symmetric around the 
points $\varphi\in\{\pi/4,3\pi/4,5\pi/4,7\pi/4\}$, i.e. these quantities
are invariant under reflections along the $(1,1)$ and $(1,-1)$
direction. These directions define the symmetry axes of the problem.
In particular, for these direction the wave vectors and particle
velocities are collinear.

\begin{figure}
\centerline{\includegraphics[width=8cm]{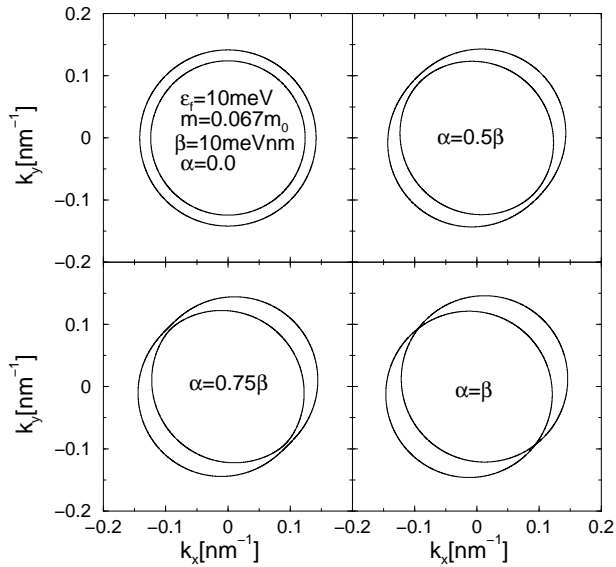}} 
\caption{Fermi contours for various values of the Rashba coefficient $\alpha$
at a Fermi energy of $10{\rm meV}$, a Dresselhaus coefficient of
$10{\rm meVnm}$, and a band mass of $0.067$ in units of the bare electron mass
$m_{0}$. The upper left panel shows the isotropic Fermi contours at
$\alpha=0$. In the upper right and lower left panel data at intermediate
values of $\alpha$ are plotted, while the lower right panel shows data
for $\alpha=\beta$. In this case the Fermi contours are two circles
having the same radius and being displaced from the origin.
\label{fig1}}
\end{figure}

The above findings for the Fermi contours are illustrated in Fig.~\ref{fig1}
where we show data for typical values for the Fermi energy, 
Dresselhaus coefficient and effective band mass
\cite{Lommer88,Jusserand92,Jusserand95} at various values for $\alpha$
\cite{Nitta97,Engels97,Heida98,Hu99,Grundler00,Sato01}. If both $\alpha$ and 
$\beta$ are nonzero the Fermi contours are anisotropic having the
$(1,1)$ and $(1,-1)$ direction as symmetry axes. 
The case $\alpha=\pm\beta$ is particular \cite{Schliemann03}. Here a new
conserved quantity given by $\Sigma:=(\sigma^{x}\mp\sigma^{y})/\sqrt{2}$
arises \cite{betanote}, 
and the spin state of the electrons becomes independent of the wave vector.
For this situation the dispersion relations are more conveniently written as
(choosing $\alpha=+\beta$)
\begin{equation}
\varepsilon_{(\pm)}(\vec k)=\frac{\hbar^{2}}{2m}
|\vec K_{(\pm)}(\vec k)|^{2}-\frac{2m\beta^{2}}{\hbar^{2}}
\label{dispersion2}
\end{equation}
where
\begin{equation}
\vec K_{(\pm)}(\vec k)=\vec k\pm\frac{\sqrt{2}m\beta}{\hbar^{2}}(1,1)
\end{equation}
is the distance vector between the centers of the circles
to points on their circumference. The double sign labeling branches
in Eq.~(\ref{dispersion2}) does {\em not} correspond to the one
in Eq.~(\ref{dispersion1}) and is therefore put in parentheses. As seen
in Fig.~\ref{fig1}, right at $\alpha=\beta$ different parts of the
dispersion branches for $\alpha\neq\beta$ merge to different circles
inducing a relabeling of branches.


\section{Boltzmann theory of anisotropic transport in two dimensions}
\label{Boltzmann}

The Boltzmann equation for transport in the two-dimensional
electron gas in the presence of fixed random impurities reads \cite{Smith89}
\begin{equation}
\frac{\partial f_{\mu}}{\partial t}
+\dot{\vec r}\cdot\frac{\partial f_{\mu}}{\partial \vec r}
+\dot{\vec k}\cdot\frac{\partial f_{\mu}}{\partial \vec k}
=\left(\frac{\partial f_{\mu}}{\partial t}\right)_{\rm coll}\,.
\end{equation}
Here $\mu$ is a band index; in the context of the previous section it 
corresponds to the double sign labeling the two dispersion branches. 
$f_{\mu}(\vec r,\vec k, t)$ 
is the usual semiclassical distribution function, and the 
collision term is given by
\begin{eqnarray}
& & \left(\frac{\partial f_{\mu}}{\partial t}\right)_{\rm coll}= \nonumber\\
& & \sum_{\mu'}\int\frac{d^{2}k'}{(2\pi)^{2}}
\Big[w(\vec k,\mu;\vec k',\mu')f_{\mu'}(\vec k')
\left(1-f_{\mu}(\vec k)\right)\nonumber\\
& & \qquad-w(\vec k',\mu';\vec k,\mu)f_{\mu}(\vec k)
\left(1-f_{\mu'}(\vec k')\right)\Big]\,,
\label{Boltzmann1}
\end{eqnarray}
where $w(\vec k,\mu;\vec k',\mu')$ is a transition probability
determined by the fixed impurities.
The semiclassical equations of motion read
\begin{equation}
\dot{\vec r}=\vec v_{\mu}(\vec k)=
\frac{\partial\varepsilon_{\mu}(\vec k)}{\hbar\partial\vec k}\qquad,\qquad
\hbar\dot{\vec k}=-e\vec E\,,
\end{equation}
where $\varepsilon_{\mu}(\vec k)$ is the dispersion of the band $\mu$,
$(-e)=-|e|$ is the electron charge, and $\vec E$ is an external
electric field in the plane of the two-dimensional gas. Assuming a
homogeneous system in a stationary state, 
$f_{\mu}(\vec r, \vec k, t)=f_{\mu}(\vec k)$, and
elastic scattering fulfilling the microreversibility condition 
$w(\vec k,\mu;\vec k',\mu')=w(\vec k',\mu';\vec k,\mu)$,
the Boltzmann equation \cite{Smith89} becomes in lowest order in $|\vec E|$
\begin{equation}
-e\vec E\cdot\vec v_{\mu}(\vec k)
\left(-\frac{\partial f^{0}}{\partial\varepsilon}\right)
=S\left[f_{\mu}(\vec k)\right]
\label{Boltzmann2}
\end{equation}
with the scattering operator
\begin{eqnarray}
& & S\left[f_{\mu}(\vec k)\right]=\nonumber\\
& & \sum_{\mu'}\int\frac{d^{2}k'}{(2\pi)^{2}}
\left[w(\vec k,\mu;\vec k',\mu')
\left(f_{\mu}(\vec k)-f_{\mu'}(\vec k')\right)\right]\,.
\label{scattop}
\end{eqnarray}
Here $f^{0}$ is the equilibrium Fermi distribution depending only on the energy
$\varepsilon$, and the derivative in Eq.~(\ref{Boltzmann2}) has to be
evaluated at $\varepsilon=\varepsilon_{\mu}(\vec k)$.

Now let $\vartheta(\vec a)$ be the angle a given vector
$\vec a$ forms with the direction of $\vec E$, fulfilling the relations
\begin{eqnarray}
\vec E\cdot\vec a & = & Ea\cos(\vartheta(\vec a))\,,\\
\left(\vec e_{z}\times\vec E\right)\cdot\vec a
& = & Ea\sin(\vartheta(\vec a))\,,
\end{eqnarray}
where $\vec e_{z}$ is the direction perpendicular to the 
two-dimensional $(xy)$-plane. The form of the transport equation
(\ref{Boltzmann2}) suggests to study the action of the scattering operator
(\ref{scattop}) on the functions 
$f_{\mu}(\vec k)=|v_{\mu}(\vec k)|\cos[\vartheta(v_{\mu}(\vec k))]$
and $f_{\mu}(\vec k)=|v_{\mu}(\vec k)|\sin[\vartheta(v_{\mu}(\vec k))]$.
In fact, by inserting these functions into (\ref{scattop}) and expressing
the angle $\vartheta(v_{\mu'}(\vec k'))$ in terms of
$\vartheta(v_{\mu}(\vec k))$ and 
$(\vartheta(v_{\mu}(\vec k))-\vartheta(v_{\mu'}(\vec k')))$
via elementary trigonometric relations,
it is easy to show that
\begin{eqnarray}
& & S\left[\left(
\begin{array}{c}
|\vec v_{\mu}(\vec k)|\cos[\vartheta(\vec v_{\mu}(\vec k))]\\
|\vec v_{\mu}(\vec k)|\sin[\vartheta(\vec v_{\mu}(\vec k))]
\end{array}
\right)\right]=\nonumber\\
& & \left(
\begin{array}{cc}
\frac{1}{\tau_{\mu}^{\parallel}(\vec k)} & 
-\frac{1}{\tau_{\mu}^{\perp}(\vec k)}\\
\frac{1}{\tau_{\mu}^{\perp}(\vec k)} & 
\frac{1}{\tau_{\mu}^{\parallel}(\vec k)}
\end{array}
\right)
\left(
\begin{array}{c}
|\vec v_{\mu}(\vec k)|\cos[\vartheta(\vec v_{\mu}(\vec k))]\\
|\vec v_{\mu}(\vec k)|\sin[\vartheta(\vec v_{\mu}(\vec k))]
\end{array}
\right)
\end{eqnarray}
with 
\begin{eqnarray}
& & \frac{1}{\tau_{\mu}^{\parallel}(\vec k)}= 
\sum_{\mu'}\int\frac{d^{2}k'}{(2\pi)^{2}}
\Biggl[w(\vec k,\mu;\vec k',\mu')\nonumber\\
& & \quad\cdot\left(1-\frac{|\vec v_{\mu'}(\vec k')|}{|\vec v_{\mu}(\vec k)|}
\cos[\vartheta(\vec v_{\mu}(\vec k))-\vartheta(\vec v_{\mu'}(\vec k'))]
\right)\Biggr]
\,,\label{taupar1}\\
& & \frac{1}{\tau_{\mu}^{\perp}(\vec k)}= 
\sum_{\mu'}\int\frac{d^{2}k'}{(2\pi)^{2}}
\Biggl[w(\vec k,\mu;\vec k',\mu')\nonumber\\
& & \quad\cdot\frac{|\vec v_{\mu'}(\vec k')|}{|\vec v_{\mu}(\vec k)|}
\sin[\vartheta(\vec v_{\mu}(\vec k))-\vartheta(\vec v_{\mu'}(\vec k'))]
\Biggr]\,.\label{tauperp1}
\end{eqnarray}
Note that $\tau_{\mu}^{\parallel}(\vec k)$ and $\tau_{\mu}^{\perp}(\vec k)$
are independent of the common direction with respect to the angles
$\vartheta(\vec v_{\mu}(\vec k))$ and $\vartheta(\vec v_{\mu'}(\vec k'))$
in the above integrals are defined, since only differences of those
angles occur. Therefore, $\tau_{\mu}^{\parallel}(\vec k)$ and 
$\tau_{\mu}^{\perp}(\vec k)$ are independent of the direction of the
electric field $\vec E$. 

Now consider the deviation $g_{\mu}(\vec k)$ of the distribution function
$f_{\mu}(\vec k)$ from equilibrium, 
\begin{equation}
g_{\mu}(\vec k)=f_{\mu}(\vec k)-f^{0}.
\end{equation}
Making the ansatz 
\begin{eqnarray}
g_{\mu}(\vec k) & = & \left(-\frac{\partial f^{0}}{\partial\varepsilon}\right)
|\vec v_{\mu}(\vec k)|
\Big(A_{\mu}(\vec k)\cos[\vartheta(\vec v_{\mu}(\vec k))]\nonumber\\
& & \qquad+B_{\mu}(\vec k)\sin[\vartheta(\vec v_{\mu}(\vec k))]\Big)
\end{eqnarray}
with two parameters $A_{\mu}(\vec k)$, $B_{\mu}(\vec k)$
one finds from the above equations
\begin{eqnarray}
A_{\mu}(\vec k) & = & \frac{-eE\tau_{\mu}^{\parallel}(\vec k)}
{1+\left(
\frac{\tau_{\mu}^{\parallel}(\vec k)}{\tau_{\mu}^{\perp}(\vec k)}
\right)^{2}}\\
B_{\mu}(\vec k) & = & \frac{-eE\tau_{\mu}^{\perp}(\vec k)}
{1+\left(
\frac{\tau_{\mu}^{\perp}(\vec k)}{\tau_{\mu}^{\parallel}(\vec k)}
\right)^{2}}\,,
\end{eqnarray}
or  
\begin{equation}
g_{\mu}(\vec k)=g^{\parallel}_{\mu}(\vec k)+g^{\perp}_{\mu}(\vec k)
\label{g1}
\end{equation}
with
\begin{eqnarray}
g^{\parallel}_{\mu}(\vec k) & = & 
-e\left(-\frac{\partial f^{0}}{\partial\varepsilon}\right)
\frac{\tau_{\mu}^{\parallel}(\vec k)}
{1+\left(
\frac{\tau_{\mu}^{\parallel}(\vec k)}{\tau_{\mu}^{\perp}(\vec k)}
\right)^{2}}\vec E\cdot\vec v_{\mu}(\vec k)\,,\label{g2}\\
g^{\perp}_{\mu}(\vec k) & = & 
-e\left(-\frac{\partial f^{0}}{\partial\varepsilon}\right)
\frac{\tau_{\mu}^{\perp}(\vec k)}
{1+\left(
\frac{\tau_{\mu}^{\perp}(\vec k)}{\tau_{\mu}^{\parallel}(\vec k)}
\right)^{2}}\nonumber\\
 & & \quad\cdot\left(\vec e_{z}\times\vec E\right)\cdot\vec v_{\mu}(\vec k)\,.
\label{g3}
\end{eqnarray}
From this distribution function the electrical current of particles
in band $\mu$ can be obtained as
\begin{equation}
\vec j_{\mu}=-e\int\frac{d^{2}k}{(2\pi)^{2}}
\vec v_{\mu}(\vec k)g_{\mu}(\vec k)\,,
\end{equation}
and the total electrical current is given by
\begin{equation}
\vec j=\sum_{\mu}\vec j_{\mu}\,.
\end{equation}
From these relations one obtains the following conductivity tensor 
\begin{equation}
\sigma=
\left(
\begin{array}{cc}
\sigma_{xx}^{\parallel}+\sigma_{xy}^{\perp} & 
\sigma_{xy}^{\parallel}-\sigma_{xx}^{\perp} \\
\sigma_{xy}^{\parallel}+\sigma_{yy}^{\perp} & 
\sigma_{yy}^{\parallel}-\sigma_{xy}^{\perp}
\end{array}
\right)
\label{condtensor}
\end{equation}
where we have introduced the definitions
\begin{eqnarray}
\sigma_{ij}^{\parallel} & = & e^{2}\sum_{\mu}\int\frac{d^{2}k}{(2\pi)^{2}}
\left(-\frac{\partial f^{0}}{\partial\varepsilon}\right)\nonumber\\
& & \qquad\cdot\frac{\tau_{\mu}^{\parallel}(\vec k)}
{1+\left(
\frac{\tau_{\mu}^{\parallel}(\vec k)}{\tau_{\mu}^{\perp}(\vec k)}
\right)^{2}}
\left(\vec v_{\mu}(\vec k)\right)_{i}\left(\vec v_{\mu}(\vec k)\right)_{j}
\,,\label{sigmaparallel1}\\
\sigma_{ij}^{\perp} & = & e^{2}\sum_{\mu}\int\frac{d^{2}k}{(2\pi)^{2}}
\left(-\frac{\partial f^{0}}{\partial\varepsilon}\right)\nonumber\\
& & \qquad\cdot\frac{\tau_{\mu}^{\perp}(\vec k)}
{1+\left(\frac{\tau_{\mu}^{\perp}(\vec k)}{\tau_{\mu}^{\parallel}(\vec k)}
\right)^{2}}
\left(\vec v_{\mu}(\vec k)\right)_{i}\left(\vec v_{\mu}(\vec k)\right)_{j}\,.
\label{sigmaperp1}
\end{eqnarray}
Several remarks are in order:

({\it i}) For an isotropic dispersion and scattering potentials isotropic
in real space, only $\sigma_{xx}^{\parallel}=\sigma_{yy}^{\parallel}$
are different from zero, and the conductivity tensor is proportional to the
unit matrix. If additionally only one 
dispersion branch is there, the parameter $\tau^{\parallel}$ becomes
\begin{eqnarray}
& & \frac{1}{\tau^{\parallel}(\vec k)}=\nonumber\\
& & \int\frac{d^{2}k'}{(2\pi)^{2}}
\Biggl[w(\vec k;\vec k')
\left(1-\cos[\vartheta(\vec k)-\vartheta(\vec k')]
\right)\Biggr]\,.
\end{eqnarray}
This is just the usual expression for the relaxation time in the isotropic
standard case \cite{Smith89} and is independent of the wave vector $\vec k$,
$\tau^{\parallel}(\vec k)=:\tau_{0}$. In fact, the above
considerations can be seen as a generalization of the standard isotropic case
to general anisotropic dispersions in two dimensions. Note that, although
the parameters $\tau^{\parallel}_{\mu}$ and $\tau^{\perp}_{\mu}$
have dimension of time, this does not mean that any relaxation time
{\em approximation} has been used to treat the case of anisotropic
dispersions. In fact Eqs.~(\ref{g1})-(\ref{g3}) constitute an 
{\em exact solution}
of the Boltzmann equation (\ref{Boltzmann2}).

({\it ii}) If in addition to 
$\sigma_{xx}^{\parallel}=\sigma_{yy}^{\parallel}\neq 0$,
the contributions $\sigma_{xy}^{\parallel}=\sigma_{yx}^{\parallel}$
are nonzero, the degeneracy of the conductivity eigenvalues is lifted.
In the case $\sigma_{xx}^{\parallel}=\sigma_{yy}^{\parallel}$ these
eigenvalues are then given by 
$\sigma_{xx}^{\parallel}\pm\sigma_{xy}^{\parallel}$
with the eigendirections $(1,\pm1)$.

({\it iii}) Provided that  $\sigma_{xx}^{\perp}=\sigma_{yy}^{\perp}\neq 0$, 
this contribution to the conductivity tensor corresponds to the anomalous or
extraordinary Hall effect. This is an antisymmetric contribution to
the conductivity tensor which does not stem from an external magnetic field
but entirely from scattering processes.
For such a contribution to be present, time reversal symmetry has to be broken.

For the case of anisotropic dispersions induced by spin-orbit coupling
as discussed in detail in the next section, we will see
that there is no anomalous Hall effect (since time reversal symmetry is
intact), but there is a symmetric off-diagonal contribution to the
conductivity tensor which stems from both
$\sigma_{xy}^{\parallel}=\sigma_{yx}^{\parallel}\neq 0$
and $\sigma_{xx}^{\perp}=-\sigma_{yy}^{\perp}\neq 0$


\section{Conductivity in the presence of spin-orbit coupling}
\label{results}

We now proceed with calculating transport properties for Fermi liquid 
electrons in two dimensions in the presence of spin-orbit coupling,
using the formalism of the previous section. To be specific, we will
evaluate the transition probabilities in the scattering operator 
(\ref{scattop}) by Fermi's golden rule,
\begin{eqnarray}
w(\vec k,\mu;\vec k',\mu') & = & \frac{2\pi}{\hbar}\frac{\nu}{A}
|(\vec k,\mu|V|\vec k',\mu')|^{2}\nonumber\\
 & & \quad\cdot
\delta\left(\varepsilon_{\mu}(\vec k)-\varepsilon_{\mu'}(\vec k')\right)\,,
\label{goldenrule}
\end{eqnarray}
where $V$ is the operator of a single scatterer and $\nu$ is the density
of scatterers.
The momentum eigenstates involved above are normalized as
\begin{equation}
(\vec k,\mu|\vec k',\mu')
=A\delta_{\vec k,\vec k'}\delta_{\mu,\mu'}
\label{norm}
\end{equation}
with $A$ being the area of the system. As a further simplification we will
consider fixed impurities with $\delta$-function shaped scattering potentials
($s$-wave approximation),
\begin{equation}
V(\vec r)=\kappa\delta(\vec r)\,,
\end{equation}
where $\kappa$ parametrizes the strength of the potential.
The square moduli of the matrix elements read
\begin{equation}
|(\vec k,\mu|V|\vec k',\mu')|^{2}
=\frac{\kappa^{2}}{2}\left(1+\mu\mu'\cos[\chi(\vec k)-\chi(\vec k')]\right).
\end{equation}
Moreover, we will concentrate on the case of zero temperature, where
the derivative of the Fermi function as it arises in the integral
expressions for transport parameters is equal to the negative of a
delta function peaked at the Fermi energy. Thus the integrations over momentum
space in Eqs.~(\ref{sigmaparallel1}), (\ref{sigmaperp1}) reduce to
integrations over the Fermi contour.

However, even with these simplifications, the integrations involved are
in general non-elementary. In order to make analytical progress
we concentrate on the case of finite Dresselhaus coupling and small
Rashba coupling ($|\alpha|\ll\beta$, section \ref{smallalpha}) and the
particular case $\alpha=\beta$ (section \ref{aeqb}).

\subsection{The case $\alpha\ll\beta$}
\label{smallalpha}

It is straightforward to expand the quantities entering the transport
parameters and conductivities discussed in section \ref{Boltzmann}
for $\alpha\ll\beta$ in lowest order in $\alpha$. However, since the
calculations are somewhat lengthy, details are given in the appendix.
The full result for the elements of the conductivity tensor 
(\ref{condtensor}) up to linear order in $\alpha$ but general
values for Dresselhaus coefficient $\beta$ and positive Fermi energy
$\varepsilon_{f}$ is stated in 
Eqs.~(\ref{fullsigmaxx}) and (\ref{fullsigmaxy}).
These expressions are still somewhat complicated but simplify
significantly if one additionally assumes that the ``Dresselhaus energy''
$\varepsilon_{D}:=m\beta^{2}/\hbar^{2}$ is small compared to $\varepsilon_{f}$
as it is usually the case for realistic situations. In other words,
defining the ``Rashba energy'' as $\varepsilon_{R}:=m\alpha^{2}/\hbar^{2}$,
we consider the situation
\begin{equation}
\varepsilon_{R}\ll\varepsilon_{D}\ll\varepsilon_{f}\,,
\end{equation}
where the Fermi energy is related to the electron density $n$ via
(cf. Eq.~(\ref{density}))
\begin{equation}
n=\frac{1}{2\pi}\left(\frac{2m}{\hbar^{2}}\varepsilon_{f}
+2\left(\frac{m}{\hbar^{2}}\right)^{2}
\beta^{2}\right)+{\cal O}\left(\alpha^{2}\right)\,.
\label{densitysmallalpha}
\end{equation}
Then one has 
\begin{eqnarray}
\sigma_{xx}=\sigma_{yy} & = & \sigma_{0}
+{\cal O}\left(\frac{\varepsilon_{R}}{\varepsilon_{f}},
\frac{\varepsilon_{D}}{\varepsilon_{f}}\right)
\label{sigmaxx}\\
\sigma_{xy}=\sigma_{yx} & = & 
\sigma_{0}(-{\rm sign}(\alpha))\frac{7}{8}
\frac{\sqrt{\varepsilon_{R}\varepsilon_{D}}}{\varepsilon_{f}}
\nonumber\\
 & & +{\cal O}\left(\frac{\varepsilon_{R}}{\varepsilon_{f}},
\sqrt{\frac{\varepsilon_{R}}{\varepsilon_{f}}}
\frac{\varepsilon_{D}}{\varepsilon_{f}}\right)
\label{sigmaxy}
\end{eqnarray}
where $\sigma_{0}$ is the usual Drude conductivity,
\begin{equation}
\sigma_{0}=\frac{e^{2}\tau_{0}n_{0}}{m}\,,
\end{equation}
and
\begin{equation}
\tau_{0}=\frac{\hbar^{3}}{\nu\kappa^{2}m}\,,
\label{taunull}
\end{equation}
\begin{equation}
n_{0}=\frac{k_{f}^{2}}{2\pi}
\end{equation}
are the momentum relaxation time and particle density, respectively, in the
absence of spin-orbit coupling. The eigenvalues of the conductivity tensor are
\begin{equation}
\sigma^{+}=\sigma_{xx}+\sigma_{xy}\quad,\quad
\sigma^{-}=\sigma_{xx}-\sigma_{xy}
\end{equation}
with corresponding eigendirections $(1,1)$ and $(1,-1)$, respectively.
These directions are the symmetry axes of the underlying dispersion
relations; the same eigendirections are found from
Eqs.~(\ref{fullsigmaxx}), (\ref{fullsigmaxy}) where the Dresselhaus energy
has not been assumed to be small compared to the Fermi energy.
From Eqs.~(\ref{sigmaxx}), (\ref{sigmaxy}), the conductivity anisotropy
$\Delta\sigma$ is given by
\begin{eqnarray}
\Delta\sigma & := & \frac{|\sigma^{+}-\sigma^{-}|}{\sigma^{+}+\sigma^{-}}
\nonumber\\
 & = & \frac{7}{8}
\frac{\sqrt{\varepsilon_{R}\varepsilon_{D}}}{\varepsilon_{f}}
+{\cal O}\left(\frac{\varepsilon_{R}}{\varepsilon_{f}},
\frac{\varepsilon_{D}}{\varepsilon_{f}}\right)
\end{eqnarray}
We note that changing the sign of $\alpha$ (by reversing the potential gradient
across the the quantum well) results in a shift by $\pi/2$ in the 
wave vector dependence of dispersion relations and eigenstates. 
Such a shift leads to a sign change in $(\sigma^{+}-\sigma^{-})$.
Therefore, this quantity contains only odd powers of $\alpha$.

The resistivity tensor $\rho$ is the inverse of the conductivity tensor 
fulfilling the relation $\vec E=\rho\vec j$. From Eqs.~(\ref{sigmaxx}),
(\ref{sigmaxy}) one finds its components to be
\begin{eqnarray}
\rho_{xx}=\rho_{yy} & = & \rho_{0}
+{\cal O}\left(\frac{\varepsilon_{R}}{\varepsilon_{f}},
\frac{\varepsilon_{D}}{\varepsilon_{f}}\right)\\
\rho_{xy}=\rho_{yx} & = & 
\rho_{0}{\rm sign}(\alpha)\frac{7}{8}
\frac{\sqrt{\varepsilon_{R}\varepsilon_{D}}}{\varepsilon_{f}}
\nonumber\\
 & & +{\cal O}\left(\frac{\varepsilon_{R}}{\varepsilon_{f}},
\sqrt{\frac{\varepsilon_{R}}{\varepsilon_{f}}}
\frac{\varepsilon_{D}}{\varepsilon_{f}}\right)
\end{eqnarray}
with $\rho_{0}=1/\sigma_{0}$. Thus, a convenient way to experimentally detect
the conductivity anisotropy is a Hall-type measurement feeding a current
in, say, the $x$-direction of the quantum well, i.e.
$j_{x}=\sigma_{xx}E_{x}$. In the absence of Rashba
coupling no voltage perpendicular to the current is generated, $E_{y}=0$,
$E_{x}=\rho_{xx}j_{x}$.
If the Rashba coupling is switched on, the off-diagonal elements
of the resistivity tensor become non-zero, and a finite transverse field
$E_{y}=\rho_{xy}j_{x}=E_{x}(\rho_{xy}/\rho_{xx})$ 
occurs. We note that this effect is similar to the
usual Hall effect with the difference that the conductivity tensor (and in turn
the resistivity tensor) is symmetric and not antisymmetric.

\subsection{The case $\alpha=\beta$}
\label{aeqb}

As discussed already in detail in Ref.~\cite{Schliemann03} and in
section \ref{dispersion}, the case $\alpha=\beta$ is special
under several aspects. Here the transport quantities are readily
obtained using the form (\ref{dispersion2}) for the dispersion relations.
As a result, the conductivity tensor is isotropic with
\begin{equation}
\sigma_{xx}=\sigma_{yy}=\sigma_{0}=\frac{e^{2}\tau_{0}n_{0}}{m}
\end{equation}
where $\tau_{0}=\hbar^{3}/\nu\kappa^{2}m$ as in Eq.~(\ref{taunull}),
and 
\begin{equation}
n_{0}=\frac{|\vec K_{f}|^{2}}{2\pi}=\frac{1}{2\pi}
\frac{2m}{\hbar^{2}}\left(\varepsilon_{f}+\frac{2m\beta^{2}}{\hbar^{2}}\right)
\end{equation}
is the density of electrons. 
At small deviations from the point $\alpha=\beta$
one should expect the conductivity tensor to develop again an anisotropy. 
However, this cannot be analyzed in the same way as the case $|\alpha|\ll\beta$
since the particle velocities and other quantities entering the
integrands in Eqs.~(\ref{taupar1}), (\ref{tauperp1}) do not allow for
an expansion in $|\alpha-\beta|$ around $\alpha=\beta$  for wave vectors
with $k_{x}=k_{y}$. At these points the dispersion branches (\ref{dispersion1})
continuously merge into two new circles when approaching $\alpha=\beta$,
cf. the lower right panel of Fig.~\ref{fig1}. Therefore, in order to evaluate
the conductivity tensor  around $\alpha=\beta$, one should use other
methods rather than expanding the dispersion relations. 
For our purposes here, we shall be content with the statement that
the conductivity tensor is of course continuous around $\alpha=\beta$,
and is isotropic exactly at that point.

\subsection{The influence of trilinear contributions to the Dresselhaus term}
\label{trilinear}

The Hamiltonian (\ref{dressel}) is derived from the bulk Dresselhaus 
spin-orbit coupling being trilinear in the momentum operators
\cite{Dresselhaus55},
\begin{eqnarray}
{\cal H}_{D}^{bulk}=\frac{\gamma}{\hbar} &\Bigl( &
\sigma^{x}p_{x}\left(p_{y}^{2}-p_{z}^{2}\right)
+\sigma^{y}p_{y}\left(p_{z}^{2}-p_{x}^{2}\right)\nonumber\\
& + & \sigma^{z}p_{z}\left(p_{x}^{2}-p_{y}^{2}\right)\Bigr)
\end{eqnarray}
with a coupling parameter $\gamma$. In a sufficiently narrow quantum well 
grown along the [001] direction 
one can approximate the operators $p_{z}$ and $p^{2}_{z}$
by their expectation values 
$\langle p_{z}\rangle$, $\langle p^{2}_{z}\rangle$. 
This leads to the
following two contributions to spin orbit coupling resulting from
bulk inversion asymmetry: The Dresselhaus term (\ref{dressel}) linear in the
momenta with $\beta=\gamma\langle p^{2}_{z}\rangle$, and the trilinear
term 
\begin{equation}
{\cal H}_{D}^{(3)}=\frac{\gamma}{\hbar}
\left(\sigma^{x}p_{x}p_{y}^{2}-\sigma^{y}p_{y}p_{x}^{2}\right).
\end{equation}
Clearly the typical magnitude of ${\cal H}_{D}^{(3)}$ compared to
the linear term ${\cal H}_{D}$ is given by the ratio
of the Fermi energy $\varepsilon_{f}$ of the in-plane motion to
the kinetic energy of the quantized degree of freedom in the growth direction.
For typical values of $\varepsilon_{f}$ of about $10{\rm meV}$ 
and not too broad quantum wells this ratio
is small, and we have therefore neglected the Dresselhaus term trilinear
in the momentum components. If desired, it is straightforward to
include this term in the calculations of transport quantities, although
the procedure becomes considerably more involved and will require
numerical calculations. However, we do not expect, for the following reasons, 
that including the
trilinear Dresselhaus term but not the Rashba term will lead to anisotropic
charge transport: The Hamiltonian
\begin{equation}
{\cal H}=\frac{\vec p^{2}}{2m}+{\cal H}_{D}+{\cal H}_{D}^{(3)}
\label{trilinham}
\end{equation}
gives the following dispersions for wave vectors
$\vec k=k(\cos\varphi,\sin\varphi)$:
\begin{eqnarray}
 & & \varepsilon_{\pm}(k,\varphi)=\frac{\hbar^{2}k^{2}}{2m}\nonumber\\
 & & \pm\sqrt{\beta^{2}k^{2}-\left(1-\cos(4\varphi)\right)
\left(\frac{\hbar^{2}}{2}\beta\gamma k^{4}-\frac{\hbar^{4}}{8}\gamma^{2}k^{6}
\right)}
\end{eqnarray}
The angular variable $\varphi$ enters only terms of $\cos(4\varphi)$
which leads to Fermi contours with fourfold symmetry, differently from the
just twofold symmetry in the case of Rashba and linear Dresselhaus term.
In particular, for the Hamiltonian (\ref{trilinham}) the dispersions are 
symmetric with respect to both the axes pairs $(1,0)$, $(0,1)$
and $(1,1)$, $(1,-1)$, and these axes pairs are the possible candidates for
eigendirections of the conductivity tensor. However, since directions in the
above pairs are equivalent due to the existence of the other pair of
symmetry axes,
we do not expect an anisotropy in tranport quantities.
Moreover, as seen above, such anisotropies arise from the interplay
of the Rashba and the Dresselhaus term and are tunable by external gates. 
Concerning gating the Rashba coefficient by an electric field
across the quantum well, one should keep in mind that such an operation
might effectively also alter the Dresselhaus coefficient 
$\beta=\gamma\langle p^{2}_{z}\rangle$ by changing the expectation value
$\langle p^{2}_{z}\rangle$. However, this additional effect cannot change
our principle findings concerning anisotropic transport.


\section{Summary and discussion}
\label{summary}

We have presented a theory of anisotropic transport in a two-dimensional
electron gas. The anisotropy in the electrical conductivity is induced
by the interplay between Rashba and Dresselhaus spin-orbit coupling 
in the semiconductor quantum well confining the electron gas.
The principle axes for anisotropic diffusive charge transport
are given by the symmetry axes of the single-particle dispersion relations,
which are anisotropic if both Rashba and Dresselhaus spin-orbit
interaction is present.
We have evaluated the conductivity tensor 
at zero temperature for scattering on fixed random impurities whose
potentials are modeled by delta functions. 
However, because of the anisotropic properties of the underlying
dispersions, we do not expect our results to change qualitatively if
other impurity potentials are considered. In particular, the differential
cross section for delta-function potentials is isotropic, which makes
obivious that our result is due to the spin-orbit induced effects and not
due to special properties of the scatterers.

To enable analytical progress in the evaluation of transport properties,
we have concentrated on the case of a finite Dresselhaus term
and a small Rashba term ($|\alpha|\ll\beta$), and on the case where 
the Rashba and Dresselhaus coefficients are equal ($\alpha=\beta$).
For $|\alpha|\ll\beta$ we have found the anisotropic corrections
to the conductivity tensor due to the presence of the Rashba term.
These findings demonstrate the principle result that diffusive
charge transport becomes anisotropic if both Rashba and Dresselhaus spin-orbit
coupling are present. This anisotropy can be tuned by external gates which
provides the possibility of detecting and investigating spin-orbit
interaction by measuring spin-unpolarized diffusive electrical currents.
Apart from possible device applications of this effect, 
the experimental observation
of such a tunable anisotropy in spin-unpolarized diffusive transport
would certainly significantly confirm and deepen our understanding
of spin-orbit coupling in semiconductors.

In our calculations we have concentrated on the Dresselhaus contributions
being linear in the momentum components, as it is appropriate for not
too wide quantum wells. The possible influence of the
trilinear Dresselhaus term is discussed in section \ref{trilinear}.

The case $\alpha=\beta$ is special under several aspects due
to the additional conserved quantity that arises at this point
\cite{Schliemann03}. Here the conductivity tensor is found to be isotropic.

Our approach to anisotropic transport in two dimensions is based on an
exact solution of the Boltzmann equation where the drift term is linearized in 
the in-plane electric field driving the current. This formalism can also
deal with the case of anisotropic single-particle dispersions and should be
seen as a generalization of the usual isotropic case. We expect this approach
to be also useful in the study of other transport effects such as
thermal conductivity, magnetothermal effects, and the anomalous Hall effect.

\acknowledgments{We thank J. Carlos Egues, C.~M. Marcus,
 and Roland Winkler for valuable
discussions. This work was supported by the NCCR Nanoscience, the Swiss NSF,
DARPA, and ARO.}


\appendix

\section{Calculation of transport properties at $|\alpha|\ll\beta$}

In this appendix we present details of the calculation of 
transport properties at $|\alpha|\ll\beta$ using Fermi's golden rule
(\ref{goldenrule}) in the case of vanishing temperature.

\subsection{Dispersion relations, eigenstates, and Fermi contours
at $|\alpha|\ll\beta$}

For the single-particle
energies and the phases $\chi(\vec k)$ entering the eigenvectors 
(\ref{eigenstate}) one finds
the following expansions:
\begin{eqnarray}
\varepsilon_{\pm}(\vec k) & = &
\frac{\hbar^{2}k^{2}}{2m}\pm\beta k
\sqrt{\left(1+\frac{\alpha^{2}}{\beta^{2}}\right)
+2\frac{\alpha}{\beta}
\frac{(\vec k^{T}\tilde\sigma^{x}\vec k)}{k^{2}}}\nonumber\\
 & = & \frac{\hbar^{2}k^{2}}{2m}\nonumber\\
 & & \pm\beta k
\left(1+\frac{\alpha}{\beta}\frac{(\vec k^{T}\tilde\sigma^{x}\vec k)}{k^{2}}
+{\cal O}\left(\frac{\alpha^{2}}{\beta^{2}}\right)\right)\,,
\end{eqnarray}
\begin{eqnarray}
\chi(\vec k) & = & 
\arg\left(\left(-k_{x}-\frac{\alpha}{\beta}k_{y}\right)
+i\left(k_{y}+\frac{\alpha}{\beta}k_{x}\right)\right)\nonumber\\
 & = & \arg\left(\left(-k_{x}+ik_{y}\right)
\left(1-\frac{\alpha}{\beta}\frac{i}{k^{2}}
\left(k_{x}+ik_{y}\right)^{2}\right)\right)\nonumber\\
 & = & \arg\Biggl(\left(-k_{x}+ik_{y}\right)\nonumber\\
 & & \cdot\left(\exp\left(-\frac{\alpha}{\beta}\frac{i}{k^{2}}
\left(k_{x}+ik_{y}\right)^{2}\right)
+{\cal O}\left(\frac{\alpha^{2}}{\beta^{2}}\right)\right)\Biggr)\nonumber\\
 & = & \arg\left(-k_{x}+ik_{y}\right)
-\frac{\alpha}{\beta}\frac{(\vec k^{T}\tilde\sigma^{z}\vec k)}{k^{2}}
+{\cal O}\left(\frac{\alpha^{2}}{\beta^{2}}\right)\,.
\label{chiexpand}
\end{eqnarray}
Here $\tilde\sigma^{x}$ and $\tilde\sigma^{z}$ are again usual Pauli
matrices acting on the two-component vectors $\vec k$.
Note that $\chi(\vec k)$ and therefore the eigenstates remain unchanged in 
if $\vec k$ points along the directions
$(1,1)$ or $(1,-1)$. This can also be seen directly from Eq.~(\ref{chi}).
The expansion of the particle velocities reads
\begin{eqnarray}
\vec v_{\pm}(\vec k) & = & \frac{\hbar\vec k}{m}\pm\frac{\beta}{\hbar}
\Biggl(\frac{\vec k}{k}-\frac{\alpha}{\beta}
\left(\frac{(\vec k^{T}\tilde\sigma^{x}\vec k)}{k^{2}}
-2\tilde\sigma^{x}\right)
\frac{\vec k}{k}\nonumber\\
 & & \quad+{\cal O}\left(\frac{\alpha^{2}}{\beta^{2}}\right)\Biggr)\,,
\end{eqnarray}
\begin{equation}
|\vec v_{\pm}(\vec k)|=\frac{\hbar k}{m}\pm\frac{\beta}{\hbar}
\pm\frac{\alpha}{\hbar}\frac{(\vec k^{T}\tilde\sigma^{x}\vec k)}{k^{2}}
+{\cal O}\left(\alpha^{2}\right)\,.
\label{velmodexp}
\end{equation}
In order to evaluate the transport parameters $\tau^{\parallel}_{\pm}$
and $\tau^{\perp}_{\pm}$ according to Eqs.~(\ref{taupar1}), (\ref{tauperp1})
one needs angles of the type $\vartheta(\vec v_{\mu}(\vec k))$. As
already remarked in section \ref{Boltzmann}, $\tau^{\parallel}_{\pm}$
and $\tau^{\perp}_{\pm}$ are independent of the direction with respect
to which these angles are defined. It is convenient to choose this direction
along the $x$-axis. Then one finds analogously to Eq.~(\ref{chiexpand})
\begin{eqnarray}
\vartheta(\vec v_{\pm}(\vec k))
 & = & \arg\left(\left(\vec v_{\pm}(\vec k)\right)_{x}
+i\left(\vec v_{\pm}(\vec k)\right)_{y}\right)\nonumber\\
 & = & \vartheta(\vec k)\pm\frac{2\alpha}{\frac{\hbar^{2}k}{m}\pm\beta}
\frac{(\vec k^{T}\tilde\sigma^{z}\vec k)}{k^{2}}
+{\cal O}\left(\alpha^{2}\right)
\end{eqnarray}
For the Fermi contour as given in Eq.~(\ref{fermimod1}) for positive 
$\varepsilon_{f}\geq 0$ one has the expansion
\begin{eqnarray}
 & & k^{f}_{\pm}\left(\varphi;\varepsilon_{f}\right)
=\sqrt{\frac{2m}{\hbar^{2}}\varepsilon_{f}
+\left(\frac{m}{\hbar^{2}}\right)^{2}\beta^{2}}\mp\frac{m}{\hbar^{2}}\beta
\nonumber\\
 & & +\frac{m\alpha}{\hbar^{2}}
\left(\frac{1}
{\sqrt{1+2\frac{\hbar^{2}}{m\beta^{2}}\varepsilon_{f}}}\mp1\right)
\sin\left(2\varphi\right)
+{\cal O}\left(\alpha^{2}\right)\,.
\label{fermimodexp}
\end{eqnarray}
Note that, at a given electron density $n$, the Fermi energy $\varepsilon_{f}$
is according to Eq.~(\ref{densitysmallalpha}) unchanged in first order
in $\alpha$. 
Moreover, when inserting the Fermi momentum in zeroth order in $\alpha$,
\begin{equation}
\left(k^{f}_{\pm}\right)_{0}
=\sqrt{\frac{2m}{\hbar^{2}}\varepsilon_{f}
+\left(\frac{m}{\hbar^{2}}\right)^{2}\beta^{2}}\mp\frac{m}{\hbar^{2}}\beta\,,
\end{equation}
in (\ref{velmodexp})
one obtains
\begin{equation}
\left(|\vec v_{\pm}(\vec k^{f}_{\pm})|\right)_{0}
=\sqrt{\frac{2\varepsilon_{f}}{m}
+\left(\frac{\beta}{\hbar}\right)^{2}}\,,
\label{modvzero}
\end{equation}
i.e. the Fermi velocity is in zeroth order in $\alpha$ independent of
the band index $\mu\in\{+,-\}$.
Using Eq.~(\ref{modvzero}), the expansion (\ref{fermimodexp})
for the Fermi momentum can be rewritten up to linear order in $\alpha$ as
\begin{eqnarray}
 & & k^{f}_{\pm}\left(\varphi;\varepsilon_{f}\right)
=\frac{m}{\hbar}
\left(\left(|\vec v_{\pm}(\vec k^{f}_{\pm})|\right)_{0}
\mp\frac{\beta}{\hbar}\right)\nonumber\\
 & & \quad\left(1\mp\frac{\alpha}{\hbar
\left(|\vec v_{\pm}(\vec k^{f}_{\pm})|\right)_{0}}
\sin(2\varphi)\right)
+{\cal O}\left(\alpha^{2}\right)\,.
\end{eqnarray}

\subsection{Transport quantities}

Using the expansions given in the previous section it is a little tedious
but straightforward to obtain expressions for the transport quantities
discussed in section \ref{Boltzmann} in up to linear order in the
Rashba coefficient $\alpha$. For the parameter 
$1/\tau_{\mu}^{\parallel}(\vec k)$, $\mu\in\{+,-\}$,
one finds in zeroth order in $\alpha$
\begin{equation}
\left(\frac{1}{\tau_{\mu}^{\parallel}(\vec k)}\right)_{0}=
\nu\kappa^{2}\frac{m}{\hbar^{3}}
\left(1+\frac{\mu}{2}\frac{\beta}{\hbar\left(|\vec v_{\mu}(\vec k)|\right)_{0}}
\right)\,,
\label{tauparzero}
\end{equation}
and the first order is given by
\begin{eqnarray}
 & & \left(\frac{1}{\tau_{\mu}^{\parallel}(\vec k)}\right)_{1}=\nonumber\\
& & -\nu\kappa^{2}\frac{m}{\hbar^{3}}
\frac{1}{2}\frac{\alpha\beta}
{\left(\hbar\left(|\vec v_{\mu}(\vec k)|\right)_{0}\right)^{2}}
\frac{(\vec k^{T}\tilde\sigma^{x}\vec k)}{k^{2}}\,.
\label{tauparone}
\end{eqnarray}
For $1/\tau_{\mu}^{\perp}(\vec k)$, the zeroth order in $\alpha$
vanishes while the first order reads 
\begin{eqnarray}
& & \left(\frac{1}{\tau_{\mu}^{\perp}(\vec k)}\right)_{1}=
\nu\kappa^{2}\frac{m}{\hbar^{3}}
\frac{\alpha}{\hbar\left(|\vec v_{\mu}(\vec k)|\right)_{0}}\nonumber\\
 & & \left(\frac{1}{2}\mu
-\frac{\beta}{\hbar\left(|\vec v_{\mu}(\vec k)|\right)_{0}}\right)
\frac{(\vec k^{T}\tilde\sigma^{z}\vec k)}{k^{2}}\,.
\label{tauperpone}
\end{eqnarray}
We now turn to the parameters entering the conductivity tensor. For the
diagonal elements of $\sigma^{\parallel}$ we find in zeroth order
\begin{eqnarray}
 & & \left(\sigma_{xx}^{\parallel}\right)_{0}
=\left(\sigma_{yy}^{\parallel}\right)_{0}=\nonumber\\
& & \frac{e^{2}}{h}
\frac{\hbar^{3}}{\nu\kappa^{2}m}\frac{m}{\hbar}
\left(\frac{2\varepsilon_{f}}{m}
+\left(\frac{\beta}{\hbar}\right)^{2}\right)
\frac
{\frac{2\varepsilon_{f}}{m}+\frac{3}{2}\left(\frac{\beta}{\hbar}\right)^{2}}
{\frac{2\varepsilon_{f}}{m}+\frac{3}{4}\left(\frac{\beta}{\hbar}\right)^{2}}
\,,
\end{eqnarray}
while the contributions in first order in $\alpha$ vanish.
The off-diagonal elements of $\sigma^{\parallel}$ are zero at $\alpha=0$,
and the first order reads
\begin{eqnarray}
& & \left(\sigma_{xy}^{\parallel}\right)_{1}
=\left(\sigma_{yx}^{\parallel}\right)_{1}=
\frac{e^{2}}{h}
\frac{\hbar^{3}}{\nu\kappa^{2}m}\frac{m}{\hbar}\frac{\alpha}{\hbar}\nonumber\\
 & & \cdot\frac{\beta}{\hbar}
\left(\frac{2\varepsilon_{f}}{m}
+\left(\frac{\beta}{\hbar}\right)^{2}\right)
\frac
{-\frac{11}{4}\frac{2\varepsilon_{f}}{m}
-\frac{27}{16}\left(\frac{\beta}{\hbar}\right)^{2}}
{\left(\frac{2\varepsilon_{f}}{m}
+\frac{3}{4}\left(\frac{\beta}{\hbar}\right)^{2}\right)^{2}}\,.
\end{eqnarray}
Finally, the off-diagonal elements of $\sigma^{\perp}$ vanish up
to linear order in $\alpha$,
\begin{equation}
\sigma^{\perp}_{xy}=\sigma^{\perp}_{yx}=0+{\cal O}\left(\alpha^{2}\right)\,.
\end{equation}
The diagonal elements are also zero at vanishing $\alpha$, and the contribution
in first order in $\alpha$ is
\begin{eqnarray}
\left(\sigma_{xx}^{\perp}\right)_{1}
=-\left(\sigma_{yy}^{\perp}\right)_{1} & = &  
-\frac{e^{2}}{h}
\frac{\hbar^{3}}{\nu\kappa^{2}m}\frac{m}{\hbar}\frac{\alpha}{\hbar}
\frac{\beta}{\hbar}
\left(\frac{2\varepsilon_{f}}{m}
+\left(\frac{\beta}{\hbar}\right)^{2}\right)\nonumber\\
 & & \cdot\frac
{\frac{2\varepsilon_{f}}{m}+\frac{27}{16}\left(\frac{\beta}{\hbar}\right)^{2}}
{\left(\frac{2\varepsilon_{f}}{m}
+\frac{3}{4}\left(\frac{\beta}{\hbar}\right)^{2}\right)^{2}}\,.
\end{eqnarray}
From this one finds the elements of the conductivity
tensor as
\begin{eqnarray}
 & & \sigma_{xx}=\sigma_{yy}=\sigma_{xx}^{\parallel}=\nonumber\\
 & & \quad\frac{e^{2}}{h}
\frac{\hbar^{3}}{\nu\kappa^{2}m}\frac{m}{\hbar}
\left(\frac{2\varepsilon_{f}}{m}
+\left(\frac{\beta}{\hbar}\right)^{2}\right)
\frac
{\frac{2\varepsilon_{f}}{m}+\frac{3}{2}\left(\frac{\beta}{\hbar}\right)^{2}}
{\frac{2\varepsilon_{f}}{m}+\frac{3}{4}\left(\frac{\beta}{\hbar}\right)^{2}}
\nonumber\\
 & & \quad+{\cal O}\left(\alpha^{2}\right)
\,,\label{fullsigmaxx}\\
 & & \sigma_{xy}=\sigma_{yx}
=\sigma_{xy}^{\parallel} 
-\sigma_{xx}^{\perp}=
\quad\frac{e^{2}}{h}
\frac{\hbar^{3}}{\nu\kappa^{2}m}\frac{m}{\hbar}\frac{\alpha}{\hbar}\nonumber\\
 & & \cdot\left(-\frac{7}{4}\right)\frac{\beta}{\hbar}
\left(\frac{2\varepsilon_{f}}{m}
+\left(\frac{\beta}{\hbar}\right)^{2}\right)
\frac
{\frac{2\varepsilon_{f}}{m}}
{\left(\frac{2\varepsilon_{f}}{m}
+\frac{3}{4}\left(\frac{\beta}{\hbar}\right)^{2}\right)^{2}}
\nonumber\\
 & & \quad+{\cal O}\left(\alpha^{2}\right)
\label{fullsigmaxy}\,.
\end{eqnarray}
Therefore, the eigenvalues of the conductivity tensor are
\begin{equation}
\sigma^{+}=\sigma_{xx}+\sigma_{xy}\quad,\quad
\sigma^{-}=\sigma_{xx}-\sigma_{xy}
\end{equation}
with corresponding eigendirections $(1,1)$ and $(1,-1)$, respectively.
These directions are the symmetry axes of the underlying dispersion relations.



%


\begin{references}

\bibitem{Wolf01}
S.~A. Wolf, D.~D. Awschalom, R.~A. Buhrman, J.~M. Daughton, S. von Molnar,
 M.~L. Roukes, A.~Y. Chtchelkanova, and D.~M. Treger, 
Science {\bf 294}, 1488 (2001).

\bibitem{Awschalom02}
{\em Semiconductor Spintronics and Quantum Computation}, eds.
D.~D. Awschalom, D. Loss, and N. Samarth, Springer, Berlin, 2002.

\bibitem{Rashba60}
E.~I. Rashba, Fiz. Tverd. Tela (Leningrad) {\bf 2}, 1224 (1960)
(Sov. Phys. Solid State {\bf 2}, 1109 (1960)); 
Y.~A. Bychkov and E.~I. Rashba, J. Phys. C {\bf 17}, 6039 (1984).

\bibitem{Dresselhaus55}
G. Dresselhaus, Phys. Rev. {\bf 100}, 580 (1955).  

\bibitem{Dyakonov86}
M.~I. Dyakonov and V.~Y. Kachorovskii,
Sov. Phys. Semicond. {\bf 20}, 110 (1986).

\bibitem{Bastard92}
G. Bastard and R. Ferreira, Surf. Science {\bf 267}, 335 (1992).

\bibitem{Das90}
B. Das, S. Datta, and R. Reifenberger, Phys. Rev. B {\bf 41}, 8278 (1990).

\bibitem{Andrada97}
E.~A. de Andrada e Silva. G.~C. La Rocca, and F. Bassani,
Phys. Rev. B {\bf 55}, 16293 (1997).

\bibitem{Voskoboynikov01}
O. Voskoboynikov, C.~P. Lee, and O. Tretyak, 
Phys. Rev. B {\bf 63}, 165306 (2001).

\bibitem{Kainz03}
J. Kainz, U. R\"ossler, and R. Winkler, cond-mat/0304017.

\bibitem{Averkiev99}
N.~S. Averkiev and L.~E. Golub, Phys. Rev. B {\bf 60}, 15582 (1999).

\bibitem{Golub00}
L.~E. Golub, N.~S. Averkiev, and M. Wilander, 
Nanotechnology {\bf 11}, 215 (2000).

\bibitem{Kiselev00}
A.~A. Kiselev and K.~W. Kim, phys. stat. sol. (b) {\bf 221}, 491 (2000).

\bibitem{Schliemann03}
J. Schliemann, J.~C. Egues, and D. Loss, 
Phys. Rev. Lett. {\bf 90}, 146801 (2003).

\bibitem{Saikin02}
S. Saikin, M. Shen, M.-C. Cheng, and V. Privman, cond-mat/0212610;
M. Shen, S. Saikin, M.-C. Cheng, and V. Privman, cond-mat/0302395.

\bibitem{Lusakowski03}
A. Lusakowski, J. Wrobel, and T. Dietl, cond-mat/0304318.

\bibitem{Winkler03}
R. Winkler, cond-mat/0305315.

\bibitem{Tarasenko02}
S.~A. Tarasenko and N.~S. Averkiev, JETP Lett. {\bf 75}, 552 (2002).

\bibitem{Pikus95} 
F.~G. Pikus and G.~E. Pikus, Phys. Rev. B {\bf 51}, 16928 (1995).

\bibitem{Miller03}
J.~B. Miller, D.~M. Zumbuhl, C.~M. Marcus, Y.~B. Lyanda-Geller, 
D. Goldhaber-Gordon, K. Campman, and A.~C. Gossard,
Phys. Rev. Lett. {\bf 90}, 076807 (2003).

\bibitem{Aleiner01}
I.~L. Aleiner and V.~I. Falko, Phys. Rev. Lett. {\bf 87}, 256801 (2001).

\bibitem{Stepanenko03}
N.~E. Bonesteel, D. Stepanenko, and D.~P. DiVincenzo, 
Phys. Rev. Lett. {\bf 87}, 207901 (2001);
G. Burkard and D. Loss, Phys. Rev. Lett. {\bf 88}, 047903 (2002);
D. Stepanenko, N.~E. Bonesteel, D.~P. DiVincenzo, G. Burkard, 
and D. Loss, cond-mat/0303474.

\bibitem{Datta90}
S. Datta and B. Das, Appl. Phys. Lett. {\bf 56}, 665 (1990).

\bibitem{Mishchenko03}
E.~G. Mishchenko and B.~I. Halperin, cond-mat/0303362. 

\bibitem{Ganichev03}
S.~D. Ganichev, V.~V. Bel'kov, L.~E. Golub, E.~L. Ivchenko, 
P. Schneider, S. Giglberger, J. Eroms, J. DeBoeck, G. Borghs, 
W. Wegscheider, D. Weiss, and W. Prettl, cond-mat/0306521.

\bibitem{Lommer88}
G. Lommer, F. Malcher, and U. R\"ossler,
Phys. Rev. Lett. {\bf 60}, 728 (1988).

\bibitem{Jusserand92}
B. Jusserand, R. Richards, H. Peric, and B. Etienne,
Phys. Rev. Lett. {\bf 69}, 848 (1992).

\bibitem{Jusserand95}
B. Jusserand, R. Richards, G. Allan, C. Priester, and B. Etienne,
Phys. Rev. B {\bf 51}, 4707 (1995).

\bibitem{Nitta97}
J. Nitta, T. Akazaki, H. Takayanagi, and T. Enoki, 
Phys. Rev. Lett. {\bf 78}, 1335 (1997).

\bibitem{Engels97}
G. Engels, J. Lange, T. Sch\"apers, and H. L\"uth,
Rev. B {\bf 55}, 1958 (1997).

\bibitem{Heida98}
J.~P. Heida, B.~J. van Wees, J.~J. Kuipers, T. M. Klapwijk, and G. Borghs,
Rev. B {\bf 57}, 11911 (1998).

\bibitem{Hu99}
C.-M. Hu, J.  Nitta, T.  Akazaki, H. Takayanagi, J Osaka, P. Pfeffer,
and W. Zawadzki, Phys. Rev. B {\bf 60}, 7736 (1999).

\bibitem{Grundler00}
D. Grundler, Phys. Rev. Lett. {\bf 84}, 6074 (2000).

\bibitem{Sato01} 
Y. Sato, T. Kita, S. Gozu, and S. Yamada, J. Appl. Phys. {\bf 89}, 8017 (2001).

\bibitem{betanote}
We note that the sign convention for the Dresselhaus coefficient $\beta$ 
used here differs from the one in Ref.~\cite{Schliemann03}. This does
not affect any physical content and can be seen as a different choice
of coordinate systems. The definition used here appears to be more common 
to the literature.

\bibitem{Smith89}
See e.g.
H. Smith and H.~H. Jensen, ``Transport Phenomena'', Clarendon Press 1989;
J.~M. Ziman, ``Principles of the theory of solids'', Cambridge University
Press 1972.

\end{references}
\end{document}